\documentclass[epj]{s}
\usepackage{graphics,epsf,epsfig}
\newcommand{\mathsc}[1]{\mathsf{\scriptscriptstyle #1}}
\newcommand{\ct}[2]{{\rm \, ct^{#1}_{#2}}}
\newcommand{\s}[2]{{\rm \, s^{#1}_{#2}}}
\renewcommand{\c}[2]{{\rm \, c^{#1}_{#2}}}
\newcommand{\cs}[2]{{\rm \, cs^{#1}_{#2}}}
\renewcommand{\sc}[2]{{\rm \, sc^{#1}_{#2}}}
\renewcommand{\Re}[0]{{\rm Re\,}}

\begin{document}
\title{Triple gauge vertices at one-loop level in THDM}
\author{Michal Malinsk\'{y}\inst{1} \and Ji\v{r}\'{\i} Ho\v{r}ej\v{s}\'{\i} \inst{2} 
%
}                     
%
%
\mail{malinsky@ipnp.troja.mff.cuni.cz}
\institute{IPNP, Faculty of Mathematisc and Physics, Charles University, Prague; S.I.S.S.A., Trieste \and IPNP, Faculty of Mathematics and Physics, Charles University, Prague}
\date{Received: date / Revised version: date}
%
\abstract{Renormalized triple gauge vertices (TGV) are examined within the 
two-Higgs-doublet model of electroweak interactions. Deviations of the TGV 
from their standard-model values are calculated at the one-loop level, in 
the on-shell renormalization scheme. As a consistency check, UV divergence 
cancellations anticipated on symmetry grounds are verified explicitly. 
Dependence of the TGV finite parts on the masses of possible heavy Higgs 
scalars is discussed briefly.
} 
\maketitle
%

\section{Introduction}
\label{intro}
The two-Higgs-doublet model (THDM) has been on stage in particle physics since the early days of spontaneously broken gauge theories (to the best of our knowledge, it has emerged first in the paper \cite{Lee:iz}, in connection with the problem of {\it T}-violation). THDM represents one of the simplest and most natural extensions of the electroweak standard model (SM): its Higgs sector contains an extra complex scalar doublet, in addition to the usual SM one. This means, among other things, that there are five physical scalar particles in the THDM spectrum, instead of the single SM Higgs boson. On the other hand, the doublet structure of the Higgs sector automatically guarantees validity of the tree-level relation $\rho=1$ for the familiar electroweak parameter $\rho={m^2_\mathsc{W}}/({m_\mathsc{Z}^{2} \cos^{2} \theta_\mathsc{W}})$, in complete analogy with the SM. 

Despite its conceptual simplicity, THDM can incorporate various kinds of "new physics"
beyond SM and thus it has always been of considerable phenomenological interest; a concise overview of its possible applications can be found e.g. in \cite{Sher:1988mj}. It remains quite popular at the present time, as the Higgs physics (or, more generally, the physics of electroweak symmetry breaking) represents the central issue of the high-energy experiments planned for the nearest future. Note that a part of the current popularity of the THDM is due to the fact that its Higgs sector essentially coincides with that of the minimal supersymmetric SM (MSSM), but is obviously less constrained. For some recent work on the THDM phenomenology see e.g. the papers \cite{Iltan:2001gf}, \cite{Krawczyk:2002df}, \cite{Gunion:2002zf}, \cite{Kanemura:2002vm} and  references therein; a recent review of the subject can be found in \cite{Sanchez}.

One of the interesting technical aspects of the general THDM is that it admits "non-decoupling effects" in the Higgs sector: the heavy Higgs scalars (i.e. such that $m_\mathsc{HIGGS} \gg  m_\mathsc{W}$) can not be simply integrated out in the low energy domain ($s \sim m^2_\mathsc{W}$) and may give non-negligible contributions to some scattering amplitudes. Note that  this is not the case in the MSSM, where the heavy Higgs bosons decouple in accordance with the Appelquist-Carazzone theorem \cite{Appelquist:tg} (for the corresponding MSSM analysis see \cite{Dobado:2000pw}).
The non-decoupling effects in THDM have been studied previously for the process 
 $e^+e^-\to W^+W^-$ \cite{Kanemura:1997wx} 
within an approximation corresponding to the equivalence theorem (ET) \cite{Cornwall:1974km} for longitudinal vector bosons. Here and in the forthcoming paper \cite{MalinskyHorejsi2} we pursue this theme further by performing more detailed calculations that enable one to go beyond the framework of the ET approximation. In the present paper we calculate, at the one-loop level, the THDM contributions to the triple gauge vertices (TGV). These vertex corrections play the most important role in the possible non-decoupling effects; some applications of the results presented here will be discussed in detail in \cite{MalinskyHorejsi2}. 
Some preliminary results in this direction have already appeared in \cite{Malinsky:2002mq}.

The paper is organized as follows. In section 2 a brief review of the THDM structure is given. Section 3 is devoted to kinematics, notation and some other technical prerequisites. In section 4 we specify the quantities of our main interest and sketch the method of their calculation. We display all relevant Feynman diagrams together with the corresponding analytic expressions. The proper cancellation of UV-divergences is demonstrated in subsection 4.6 . In section 5 we present a brief discussion of the results, in particular the mass dependence of the finite parts of renormalized TGV.
Most of the technical details (structure of the Higgs-vector-boson interactions, coupling constants, useful integrals) are deferred to appendices.        


\label{sec:2}

\section{Basic structure of THDM}
In this paper we adopt the `classic' notation of the book \cite{Gunion:1989we}. We do not restrict ourselves to any particular realization of the THDM Higgs potential unless stated otherwise. As we shall see, although the relevant one-loop on-shell counterterms can in principle involve contributions descending from the Higgs self-couplings dictated by the particular realization of the model, they  cancel out in the counterterm prescription.


\subsection{The Higgs potential and spectrum}

\label{subsec:2:1}

The most general THDM Higgs potential can be written as
\begin{eqnarray}
	V(\Phi_1,\Phi_2)& = &m_{11}^2\Phi_1^\dagger\Phi_1+m_{22}^2\Phi_2^\dagger\Phi_2-
        (m_{12}^2\Phi_1^\dagger\Phi_2+ {\rm h.c.})+ 
\nonumber\\
&&       + \frac{\lambda_1}{2}(\Phi_1^\dagger\Phi_1)^2+
        \frac{\lambda_2}{2}(\Phi_2^\dagger\Phi_2)^2+ 
\nonumber\\ 
&&	+
        \lambda_3(\Phi_1^\dagger\Phi_1)(\Phi_2^\dagger\Phi_2)
	+\lambda_4(\Phi_1^\dagger\Phi_2)(\Phi_2^\dagger\Phi_1)+
\nonumber \\
&&       +
       \!\!\!\!\!\!\!\!\!\!\!\!\!\!\!\!\!\!\!\!\!\!\!\!\!\!\!\!\!\!\!\!\!\! 
	+\left\{\frac{\lambda_5}{2}(\Phi_1^\dagger\Phi_2)^2
	+[\lambda_6(\Phi_1^\dagger\Phi_1)+\lambda_7(\Phi_2^\dagger\Phi_2)]
        (\Phi_1^\dagger\Phi_2)+ {\rm h.c.}\right\}
\nonumber
\end{eqnarray}
It is convenient to parametrize the doublets by means of eight real scalar fields
\[
	  \Phi_1 =
  	\frac{1}{\sqrt{2}}\left(\begin{array}{c}
    \phi_1+i\phi_2 \\ \phi_3+i\phi_4
    \end{array}\right),
    \qquad
    \Phi_2 =
    \frac{1}{\sqrt{2}}\left(\begin{array}{c}
    \phi_5+i\phi_6 \\ \phi_7+i\phi_8
    \end{array}\right)
\]
The asymmetric vacuum is chosen so that
\[
  	\langle \phi_1 \rangle = v_1, \qquad
  	\langle \phi_7 \rangle = v_2, \qquad
  	\langle \phi_i \rangle = 0 \quad {\rm for} \quad i\neq 1,7
\]
where the $v_1$ and $v_2$ are real constants. It is useful to introduce the angle $\beta$ through $\tan \beta \equiv {v_2}/{v_1}$.
\\
The mass-squared matrix $M^2_{ij}\equiv \left\langle \partial^2 V/\partial \phi_i\partial \phi_j \right\rangle$ turns out to be  block-diagonal, with each block of dimension 2.
The diagonalization is straightforward and gives rise to the  spectrum consisting of
\begin{itemize}

\item two charged states $H^\pm$ and $G^\pm$  with masses $m^2_{H^\pm}\neq 0$ and $m^2_{G^\pm} =  0$:
\begin{eqnarray}
      G^-& = &  \frac{1}{\sqrt{2}}
      [(\phi_1+i\phi_2)\cos{\beta}+(\phi_5+i\phi_6)\sin{\beta}]
      \nonumber \\
      H^-& = & \frac{1}{\sqrt{2}}
      [-(\phi_1+i\phi_2)\sin{\beta}+(\phi_5+i\phi_6)\cos{\beta}]
\nonumber
\end{eqnarray}
\item two neutral pseudoscalar states $A^0$ and $G^0$ with $m^2_{A^0}\neq 0$ and $m^2_{G^0}=0$, which are mixed through the same angle $\beta$:
\begin{eqnarray}
      G^0& = &  \phi_4\cos{\beta}+\phi_8\sin{\beta}
      \nonumber \\
      A^0& = &  -\phi_4 \sin{\beta}+\phi_8\cos{\beta}
\nonumber
\end{eqnarray}
\item two neutral scalar states $H^0$ and $h^0$ with  $m^2_{H^0} > m^2_{h^0}\neq 0$, mixed through another angle $\alpha$:
\begin{eqnarray}
      H^0& = &  (\phi_1-v_1)\cos{\alpha}+(\phi_7-v_2)\sin{\alpha}
      \nonumber \\
      h^0& = &  -(\phi_1-v_1)\sin{\alpha}+(\phi_7-v_2)\cos{\alpha}
\nonumber
\end{eqnarray}

\end{itemize}
Then the doublets $\Phi_1$ and $\Phi_2$ can be written in terms of the physical fields as
\begin{eqnarray}
	  \Phi_1 &\!\! =\!\! &
  	\frac{1}{\sqrt{2}}\left[\!\begin{array}{c}
    \sqrt{2}G^+\cos\beta - \sqrt{2}H^+\sin\beta
\\
	  H^0\cos\alpha-h^0\sin\alpha+v_1+iG^0\cos\beta-iA^0\sin\beta
    \end{array}\right] \nonumber \\
    \Phi_2 &\!\! =\!\! &
  	\frac{1}{\sqrt{2}}\left[\!\begin{array}{c}
    \sqrt{2}G^+\sin\beta + \sqrt{2}H^+\cos\beta
\\
  	H^0\sin\alpha+h^0\cos\alpha+v_2+iG^0\sin\beta+iA^0\cos\beta
    \end{array}\right]
	\nonumber
\\
&&
    \label{expansions}
\end{eqnarray}
Note that the $G^\pm$ and $G^0$ are the unphysical Goldstone bosons to be eaten in the Higgs mechanism and $h^0$, $H^0$, $A^0$ and $H^\pm$ represent the physical scalars.



\subsection{Interactions of THDM Higgses with vector bosons}

\label{subsec:2:2}

Interactions of the Higgs bosons with intermediate vector bosons descend from the gauge-invariant kinetic term
\begin{equation}
    {\cal L}^{\mathsc{KIN}}_{\mathsc{HIGGS}}=
    (D_\mu \Phi_1)^\dagger (D^\mu \Phi_1)+
    (D_\mu \Phi_2)^\dagger (D^\mu \Phi_2)
\label{lagrangian}
\end{equation}
Here the covariant derivative is defined by
\begin{eqnarray}
		D_\mu &\equiv &
		\partial_\mu +i \frac{g}{\sqrt{2}}(W^+_\mu T^+ + W^-_\mu T^-) +ieA_\mu 	Q +
\nonumber \\
 && +i \frac{g}{\cos{{\theta_\mathsc{W}}}}
		Z_\mu(T_3-Q \sin^2{{\theta_\mathsc{W}}}) 
\nonumber
\end{eqnarray}
where the weak isospin operators $T^\pm\equiv T_1\pm iT_2$ and $T_3$
are expressed in terms of the Pauli matrices as $T_i\equiv \frac{1}{2}\tau_i$ and the charge operator is given by $Q=T_3+Y_\mathsc{W}$, with $Y_\mathsc{W}$ denoting the weak hypercharge.
Inserting now the expansion (\ref{expansions}) into (\ref{lagrangian}) one gets the relevant part of the physical lagrangian; its detailed form is deferred to the Appendix \ref{AppendixA}.


\subsection{Interactions with fermions}

\label{subsec:2:3}

In general there are several different realizations of THDM \cite{Iltan:2001gf},\cite{Gunion:1989we}. 
They differ mainly in the structure of Yukawa couplings of the THDM  
Higgs bosons to fermions. 
As we shall see, the one-particle irreducible (1PI) one-loop graphs relevant for our purpose (i.e. those representing the leading-order corrections to the tree-level structure of TGV in THDM) do not include such Yukawa vertices. Therefore we need not distinguish various THDM types in the subsequent computation.



\section{Structure of triple gauge vertices}
\label{sect:3}

As in the SM case, there are two triple gauge vertices in THDM: $\gamma WW$ and $ZWW$.
Let us denote the corresponding 1PI Green functions by
$$
\label{irrvertices}
\parbox{14mm}{\epsffile{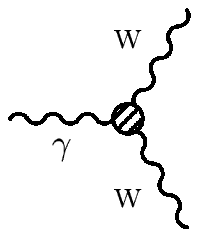}}
 \equiv -ic_{\gamma}\Gamma^\mathsc{\gamma WW}_{\sigma\mu\nu}(q_i) 
\nonumber
\quad
\parbox{15mm}{\epsffile{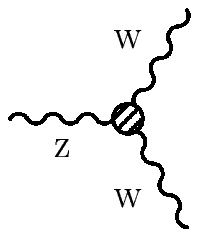}} 
 \equiv -ic_\mathsc{Z}\Gamma^\mathsc{ZWW}_{\sigma\mu\nu}(q_i) 
$$
Here $c_\gamma \equiv e$, $c_\mathsc{Z} \equiv e /\tan{}{\theta_\mathsc{W}}$ are tree-level triple gauge couplings,  $q_1$, $q_2$ and $q_3$ denote the  four-momenta of  the $W^+$, $W^-$ and $\gamma$ (or $Z$), all of them taken as {\it outgoing}.
\subsection{Tree-level triple gauge vertices}
Taking into account the momentum conservation, the {\it tree-level TGV} has the familiar form
(see e.g. \cite{Pokorski:ed})
\begin{eqnarray}
\Gamma^\mathsc{VWW}_{\sigma\mu\nu}(q_1,q_2)  & =&  
\nonumber \\
&& \!\!\!\!\!\! \!\!\!\!\!\! \!\!\!\!\!\!\!\!\!\!\!\! \!\!\!\!\!\! \!\!\!\!\!\!\!\!\!\!\!\! \!\!\!\!\!\! \!\!\!\!\!\!
		(q_1-q_2)_\sigma g_{\mu\nu}+ (2q_2+q_1)_\mu g_{\sigma\nu}-
		(2q_1+q_2)_\nu g_{\sigma\mu}
\end{eqnarray}
where $V$ stands for $\gamma$ or $Z$.
It is convenient to define
\begin{equation}\label{basicstructures1}
  C^{1}_{\sigma\mu\nu}   \equiv  {q_1}_\sigma g_{\mu\nu} \qquad
  C^{2}_{\sigma\mu\nu}   \equiv  2q_{2\mu} g_{\sigma\nu} \qquad
  C^{3}_{\sigma\mu\nu}   \equiv  q_{1\mu}  g_{\sigma\nu}
\end{equation}
In terms of these quantities we can write
\begin{equation}
\Gamma^\mathsc{VWW}_{\sigma\mu\nu}  =
C^1_{\sigma\mu\nu}+C^3_{\sigma\mu\nu}+C^3_{\sigma\mu\nu}+sym.
\equiv 
T_{\sigma\mu\nu}\label{tree-level_TGV}
\end{equation}
where '$+ sym.$' denotes the interchange $q_1 \leftrightarrow -q_2$, $\mu \leftrightarrow \nu$ in the preceding expression. 

\subsection{Triple gauge vertices at one-loop order}
\label{subsec3:2}
The full one-loop renormalized TGV receive much more involved structure. 
Let us divide the set of all relevant one-loop diagrams into two subsets, 
where $\Gamma=\Gamma_\mathsc{F}+\Gamma_\mathsc{B}$ 
represent the graphs involving one fermionic and one bosonic loop respectively. 
As we have already noted, the fermionic one-loop contributions to TGV in SM and THDM are the same and therefore there is no need to discuss them.
\\
It is clear that the on-shell TGV can only involve tensors at most trilinear in external momenta (with bilinear terms obviously absent).
Thus, we can decompose the bosonic part into a basis consisting of the linear terms (\ref{basicstructures1}) and trilinear ones
\begin{eqnarray}\label{basicstructures2}
  C^{4}_{\sigma\mu\nu}   \equiv  \frac{1}{m^2_\mathsc{W}}{q_1}_\sigma q_{1\mu} q_{1\nu} \qquad
  C^{5}_{\sigma\mu\nu}   \equiv  \frac{1}{m^2_\mathsc{W}}{q_1}_\sigma q_{1\mu} q_{2\nu} \\
  C^{6}_{\sigma\mu\nu}   \equiv  \frac{1}{m^2_\mathsc{W}}q_{1\sigma} q_{2\mu} q_{1\nu} \qquad
  C^{7}_{\sigma\mu\nu}   \equiv  \frac{1}{m^2_\mathsc{W}}q_{2\sigma} q_{1\mu} q_{1\nu} \nonumber
\end{eqnarray}
and write $\Gamma$ in the form
\begin{equation}
  \label{TGVone-loop1}
\Gamma^\mathsc{VWW}_{\sigma\mu\nu}  = \left(T_{\sigma\mu\nu}+
  \sum \Gamma^\mathsc{VWW}_{\sigma\mu\nu}+\delta Z_\mathsc{TGV}T_{\sigma\mu\nu}\right) +\Gamma^\mathsc{VWW}_\mathsc{F}
\end{equation}
The first term in the brackets is the tree-level part (\ref{tree-level_TGV}), the second one comes from the sum of all relevant bosonic one-loop 1PI graphs and the third one represents the corresponding counterterm. 
The second term can now be expanded as  
\begin{equation}
\label{exansionofsumofgraphs}
\sum \Gamma^\mathsc{VWW}_{\sigma\mu\nu}=
  \sum_{i=1}^7 \Pi_i^\mathsc{VWW}(q_1^2,q_2^2,m_j^2)C^i_{\sigma\mu\nu}+sym.	
\end{equation}
and using this we can rewrite the full one-loop renormalized Green function (\ref{TGVone-loop1}) in the form
\begin{eqnarray}
  \label{TGVone-loop2}
  \Gamma^\mathsc{VWW}_{\sigma\mu\nu}  & = & 
  \left.\sum_{i=1}^3\left(1+\delta Z_\mathsc{TGV}+ \Pi_i^\mathsc{VWW}\right)C^i_{\sigma\mu\nu}+\right.
\\
  & + & \sum_{i=4}^7 \Pi_i^\mathsc{VWW}C^i_{\sigma\mu\nu}+\Gamma^\mathsc{VWW}_\mathsc{F}+sym.
\nonumber
\end{eqnarray}
Concerning the notation, let us add that the symbol $\delta Z$ corresponds to the usual split of the renormalization constant $Z=1+\delta Z$.


\section{Deviations of THDM one-loop triple gauge vertices from SM}
\label{sect:4}
As in the SM case \cite{Bohm:1987ck} the full
one-loop corrected TGV in THDM are very complicated because of
the rich field contents of the theory. Since the models differ only
in the Higgs sector, we can utilize the previous results in
the non-Higgs sector and compute only the graphs which are not common to both models.
These additional pieces can even be used separately in many situations.
For example, it is shown in \cite{MalinskyHorejsi2} that the leading 1-loop correction to the SM value 
of differential cross-sections of $e^+e^-\to W^+W^-$ in THDM can be written in the form
\[
	  \frac{{\rm d}\sigma^\mathsc{THDM}}{{\rm d}\sigma^\mathsc{SM}}= 1+2 {\rm Re} 		
	  \frac{\Delta {\cal 	
		M}_\mathsc{1-LOOP}\left[\Delta\Gamma^\mathsc{VWW}\right]}
		{M_\mathsc{TREE}^\mathsc{SM}}+\ldots
\]
Here the structure of the term $\Delta {\cal M}_\mathsc{1-LOOP}$ is determind  by the {\it differences of the one-loop renormalized Green functions in THDM and SM} defined as
\\ \mbox{}
\begin{eqnarray}
	 \Delta\Gamma^\mathsc{VWW}_{\sigma\mu\nu} & \equiv &
	\left[\Gamma^\mathsc{VWW}_{\sigma\mu\nu}\right]_\mathsc{THDM}
	-\left[\Gamma^\mathsc{VWW}_{\sigma\mu\nu}\right]_\mathsc{SM}
\end{eqnarray}
Using (\ref{TGVone-loop2}) we can recast the last expression as
\begin{eqnarray}
  \label{TGVdifference1}
  \Delta\Gamma^\mathsc{VWW}_{\sigma\mu\nu} &  = &\sum_{i=1}^3\left(\Delta\delta Z_\mathsc{TGV}+ \Delta\Pi_i^\mathsc{VWW}\right)C^i_{\sigma\mu\nu}+ 
\\
  & & +\sum_{i=4}^7 \Delta\Pi_i^\mathsc{VWW}C^i_{\sigma\mu\nu}+sym. \nonumber	
\end{eqnarray}
(note the cancellation of the fermionic part). Here we have denoted
\begin{eqnarray}
\label{PIdifferences}
 \Delta \delta Z_\mathsc{TGV} & \equiv &(\delta Z_\mathsc{TGV})_\mathsc{THDM}-(\delta Z_\mathsc{TGV})_\mathsc{SM} \\
 \Delta \Pi_i^\mathsc{VWW} & \equiv & (\Pi_i^\mathsc{VWW})_\mathsc{THDM}-( \Pi_i^\mathsc{VWW})_\mathsc{SM} \nonumber
\end{eqnarray}
Our goal is therefore to write down the quantities $\Delta\delta Z_\mathsc{TGV}$ and $\Delta\Pi_i^\mathsc{VWW}$ for $i=1\ldots7$. 
Note that there is a nontrivial consistency check for the resulting expressions
: The divergent parts of the $\Delta\Pi_i^\mathsc{VWW}$ for $i=1,2,3$ and $V=\gamma, Z$ 
must be equal in order to be successfully 'eaten' by the divergences of 
$\Delta\delta Z_\mathsc{TGV}$. 
Moreover, the quantities 
$\Delta\Pi_i^\mathsc{VWW}$ for $i=4\ldots7$ have to be finite because 
the gauge-invariant lagrangian does not contain corresponding counterterms.



\subsection{Renormalization framework}
\label{subsect:4:1}
 In this paper we use the set of renormalization constants introduced in \cite{Pokorski:ed}
 (except that we use the symbol $\delta Z_\mathsc{TGV}$ instead of $\delta Z_g$ of \cite{Pokorski:ed} 
 to express the fact that $\delta Z_\mathsc{TGV}$ is {\it not} the usual
 counterterm corresponding to the fermion-gauge-boson vertex). 
 The parameters are fixed so that the renormalized 
 propagators have poles at the corresponding physical masses and the residues are normalized to 1. 
 As we show below, a  Ward identity connects the TGV counterterm $\delta Z_\mathsc{TGV}$ 
 to the wave-function renormalization constant  $\delta Z_\mathsc{W}$ of the $W$ boson propagator. 
 
The only information we need is the structure of the vector boson propagator counterterms 
in the on-shell ren. scheme. The one-loop renormalized inverse propagator has the general form
\begin{eqnarray}
		\Gamma_{\mu\nu}^\mathsc{VV}(k) & = & \Gamma^{(0)}_{\mu\nu}(k)+\Pi_{\mu\nu}(k)+
		\delta 
		Z_\mathsc{V}k^2 P^T_{\mu\nu}-\delta m^2_\mathsc{V} g_{\mu\nu}+ \nonumber \\
 && +{\rm gauge \,\,\, 
		dependent\,\,\,term} \nonumber 
\end{eqnarray}
where $\Gamma^{(0)}_{\mu\nu}(k)$ is the zeroth-order inverse propagator (in Feynman gauge $\Gamma^{(0)}_{\mu\nu}(k)=(k^2-m^2)(P^T_{\mu\nu}+P^L_{\mu\nu})$  ).
The on-shell counterterms are fixed by 
\begin{eqnarray}
\label{basic_definitio_Zm}
&\delta Z_\mathsc{V}  = -\left[\Pi_\mathsc{VV}^T(m^2_\mathsc{V})+
m^2_\mathsc{V}\frac{d}{dq^2}\Big|_{q^2=m_\mathsc{V}^2}{\Pi^T_\mathsc{VV}}(q^2)\right]
 \\\label{basic_definitio_m}
&\delta m_\mathsc{V}^2  =
-m_\mathsc{V}^4\frac{d}{dq^2}\Big|_{q^2=m_\mathsc{V}^2}{\Pi^T_\mathsc{VV}}(q^2) \nonumber
\end{eqnarray}
and $\Pi^T_\mathsc{VV}(q^2)$ is the coefficient of the transverse projection operator in
the decomposition of the vector boson self-energy $i\Pi_{\mu\nu}^\mathsc{VV}(k)$, namely
\[
		i\Pi_{\mu\nu}^\mathsc{VV}(k)\equiv ik^2\Pi^T_\mathsc{VV}(k^2){P}^T_{\mu\nu} + 
		ik^2\Pi^L_\mathsc{VV}(k^2)P^{L}_{\mu\nu} 
\]


\subsection{Relation $\Delta \delta Z_\mathsc{TGV}= \Delta \delta Z_\mathsc{W}$}
\label{subsect:4:2}
The photon mass counterterm is in our scheme expressed as 
\[
		\delta m_\gamma^2 = -\s{2}{\theta}\c{2}{\theta}m_\mathsc{Z}^2 
		Z_\mathsc{H}\left(1-\frac{\delta 	
		v}{v}\right)^2(1-Z_\mathsc{W}Z_\mathsc{TGV}^{-1})^2 
\]
(see Appendix C of \cite{Pokorski:ed}).
Utilizing (\ref{basic_definitio_m}) one obtains
\begin{eqnarray}
		\s{2}{\theta}\c{2}{\theta}m_\mathsc{Z}^2 Z_\mathsc{H}
		\left(1-\frac{\delta v}{v}\right)^2(1-Z_\mathsc{W} Z_\mathsc{TGV}^{-1})^2 &&
\nonumber \\	
		 = \lim_{m\to 0}m^4 
		\frac{d}{dq^2}\Big|_{q^2=m^2}{\Pi^T_\mathsc{\gamma\gamma}}(q^2)=0 &&
\nonumber
\end{eqnarray}
which yields 
$Z_\mathsc{TGV}=Z_\mathsc{W}$
or equivalently $\delta  Z_\mathsc{TGV}=\delta Z_\mathsc{W}$ and thus
\begin{equation}
\Delta\delta Z_\mathsc{TGV}=\Delta \delta Z_\mathsc{W}
\end{equation} 
This is the key relation in the following calculation.



\subsection{Computation of $\Delta\delta Z_\mathsc{W}$}
\label{subsect:4:3}
Note first that there is no tadpole contribution to the $\delta Z_\mathsc{W}$ computed by means of (\ref{basic_definitio_Zm}). This is gratifying in view of the complicated structure of the trilinear Higgs couplings.
Defining as usual
$\Delta \Pi^T_\mathsc{WW}\equiv [\Pi^T_\mathsc{WW}]_\mathsc{THDM}-[\Pi^T_\mathsc{WW}]_\mathsc{SM}$
we can write
\[
\Delta\delta Z_\mathsc{W}  = -\left[\Delta\Pi_\mathsc{WW}^T(m_\mathsc{W}^2)+
m_\mathsc{W}^2\frac{d}{dq^2}\Big|_{q^2=m_\mathsc{W}^2}{\Delta\Pi^T_\mathsc{WW}}(q^2)
\right]
\]


\paragraph{Relevant diagrams\label{graphs_for_DZww}:}

\mbox{} \\
There are only two relevant topologies contributing to 
$\Delta\delta Z_\mathsc{W}$: 
\begin{equation}
  \label{graph1}
  \epsffile{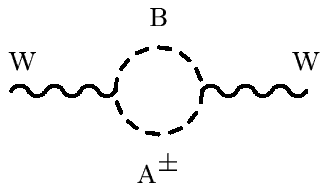}a)\qquad 
  \epsffile{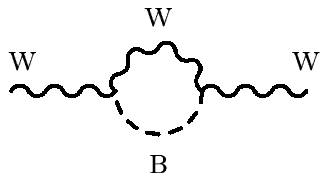}b)
\end{equation}
The two scalar lines in the first graph correspond to configurations: 
$\{\eta G^\pm\}$ in SM and $\{h^0 G^\pm$, $H^0 G^\pm$, $h^0 H^\pm$, $H^0 H^\pm$, $A^0H^\pm\}$ 
in THDM. The internal scalar and vector lines in the second case 
are 
$\{\eta W^\pm\}$ in SM and $\{h^0 W^\pm,H^0 W^\pm\}$ in THDM. 
All remaining graphs are common to both models and therefore 
cancel in the relative quantities.
\\
Using the dimensional regularisation with $d=4-2\varepsilon$ the graphs (\ref{graph1}a) give
\begin{eqnarray}\label{CT1}
		\Delta\delta Z_\mathsc{W}^a & = &  
		\left(\sum_\mathsc{THDM}-\sum_\mathsc{SM}\right)
		\left|g_\mathsc{WA^\pm B}\right|^2
		\frac{1}{16 \pi^2}\times
\nonumber\\
		&\times &\left[
		\frac{1}{3}C_\mathsc{UV}-
		2\int_0^1{\rm d}x x(1-x)
		\log \frac{D_x^\mathsc{A^\pm B}(m_\mathsc{W}^2)}{\mu^2}\right]
\nonumber
\end{eqnarray}
while the type (\ref{graph1}b) yields
\begin{equation}\label{CT2}
		\Delta\delta Z_\mathsc{W}^b = \left(\sum_\mathsc{THDM}-\sum_\mathsc{SM}\right)
		g_\mathsc{WWB}^2 \frac{1}{16\pi^2} \int_0^1{\rm d}x 
		\frac{x(1-x)}{D_x^\mathsc{WB}(m_\mathsc{W}^2)} 
\end{equation}
Here we use the abbreviations
\begin{eqnarray}
  D_x^\mathsc{XY}(q^2) & \equiv & m_\mathsc{X}^2(1-x) + m_\mathsc{Y}^2 x -q^2 x(1-x) \nonumber\\
  C_\mathsc{UV} & \equiv & \frac{1}{\varepsilon}-\gamma_E+\log 4\pi\nonumber
\end{eqnarray}
The explicit expressions for the coupling constants $g_\mathsc{WA^\pm B}$ and $g_\mathsc{WWB}$ can be found in the \ref{couplings}.

\subsection{Computation of $\Delta\Pi_i^\mathsc{VWW}$
\label{subsect:4:4}
\label{GRAPHS}}
Let us first present the list of all diagrams that are not common to both models and therefore do not cancel trivially in $\Delta\Pi_i^\mathsc{VWW}$. The charged bosons propagating in the loops are denoted by a generic symbol $A^\pm$ and the neutral ones by $B$ and $C$.
All topologies are supplemented by the list of relevant field configurations.
\paragraph{Relevant topologies :}
 \begin{displaymath}
 \epsffile{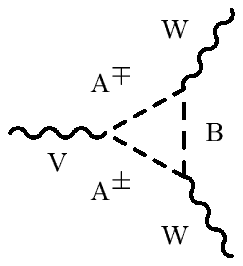}\quad 1)\,\,
 \epsffile{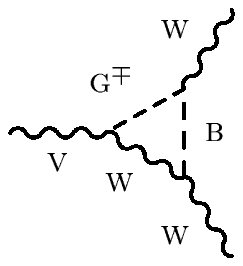}\quad 2)\,\,
 \epsffile{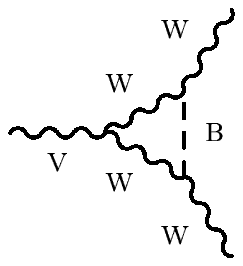}\quad 3)
\end{displaymath}
\begin{equation}\label{commontopologies}
 \epsffile{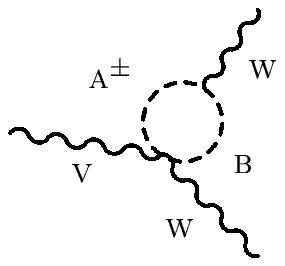}\quad 4)\quad
 \epsffile{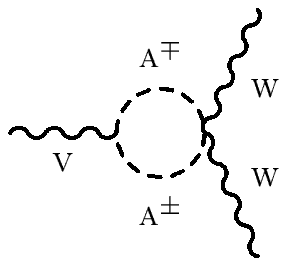}\quad 5)\quad
\end{equation}
\begin{displaymath}
 \epsffile{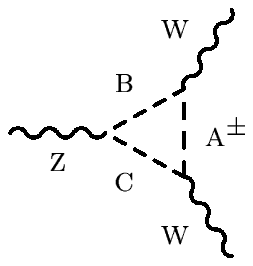}\quad 6)\quad
 \epsffile{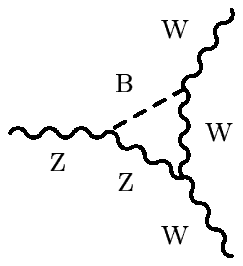}\quad 7)\quad
\end{displaymath}
Note that $V$ in $1)$ -- $5)$ corresponds to $\gamma$ and $Z$ -- these topologies are
common to both the $\Delta\Gamma^{\gamma\mathsc{WW}}$
and $\Delta\Gamma^{\mathsc{ZWW}}$ functions. 
The graphs $6)$  and $7)$ contribute
only to $\Delta\Gamma^{\mathsc{ZWW}}$.
\paragraph{Relevant field configurations:}
\mbox{}\\
The configurations of the internal lines denoted by $A^\pm$, $B$, and $C$ in the
Feynman diagrams above can be read off from the table \ref{tabulka1}.

\begin{table}
\label{tabulka1}
\caption{Field configurations in the triangular graphs (\ref{commontopologies})}
\begin{tabular}{l|l}
case & SM and THDM field configurations   \\
\noalign{\smallskip}\hline
$1)$& SM: $A^\pm B$ $=$  $G^\pm \eta$   \\
&  THDM: $A^\pm B$ $=$ $G^\pm h^0$, $G^\pm H^0$, $H^\pm A^0$, \\ 
& \hskip 1.2cm $H^\pm h^0$,$H^\pm H^0$  \\ 
$2)$& SM: $B$ $=$ $\eta$   \\
&  THDM: $B$ $=$ $h^0$, $H^0$  \\
$3)$&SM: $B$ $=$ $\eta$    \\
&  THDM: $B$ $=$ $h^0$, $H^0$   \\
$4)$& SM: $A^\pm B$ $=$  $G^\pm \eta$   \\
& THDM: $A^\pm B$ $=$  $G^\pm h^0$, $G^\pm H^0$, $H^\pm A^0$,  \\ 
& \hskip 1.2cm $H^\pm h^0$, $H^\pm H^0$   \\
$5)$&  SM: $A^\pm$ $=$ nothing   \\
&  THDM: $A^\pm$ $=$ $H^\pm$  \\
$6)$&SM: $A^\pm BC$ $=$ $G^\pm G^0\eta$   \\
&THDM: $A^\pm BC$ $=$ $G^\pm G^0h^0$, $G^\pm G^0H^0$,  \\ 
& \hskip 1.2cm $H^\pm A^0h^0$, $H^\pm A^0H^0$    \\
$7)$&SM: $B$ $=$ $\eta$    \\
&THDM: $B$ $=$ $h^0$, $H^0$   \\
\noalign{\smallskip}\hline
\end{tabular}
\end{table}

Note that according to the previous definitions the symbols 
$\Delta\Gamma^{\mathsc{VWW}(i)}$ denote the differences of the contributions coming from the previous graphs with the overall (tree) coupling constants thrown away. Thus, if we denote by $G^{(i)}$ the expressions obtained from these graphs just  by using the appropriate Feynman rules, one has  
\begin{equation}
\label{tgvbygraphs}
-ic_\mathsc{V}\Delta\Gamma^{\mathsc{VWW}(j)}
 \equiv G^{(j)}_\mathsc{THDM}-G^{(j)}_\mathsc{SM}
\end{equation}
The couplings $c_\mathsc{V}$ are defined in (\ref{irrvertices}). 
With all this at hand it is already easy to extract the corresponding 
 ${\Delta\Pi_i^\mathsc{VWW}}$ out of $\sum_j\Delta\Gamma^{\mathsc{VWW}(j)}$
in accordance with definitions (\ref{exansionofsumofgraphs}) and (\ref{PIdifferences}).

\subsection{Evaluation of ${\Delta\Pi_i^\mathsc{VWW}}$}
\label{subsect:4:5}

Let us now summarize the contributions of graphs
(\ref{commontopologies}) to the TGV differences (\ref{tgvbygraphs}). The
coupling constants and the integrals appearing in the following expressions can
be found in the appendices.
The additional numerical factors are usually due to the symmetry
properties of the graphs.
\\
The diagrams $1)$ -- $5)$ that contribute to both
 $\Delta\Gamma^\mathsc{\gamma WW}$ and $\Delta\Gamma^\mathsc{ZWW}$  
yield  the expressions
\begin{eqnarray}
		-ic_\mathsc{V} 
		{\Delta\Gamma^\mathsc{V WW}_{\sigma\mu\nu}}^{(1)}
& = &
		\left(\sum_\mathsc{THDM}-\sum_\mathsc{SM}\right)
		\frac{1}{2}g_\mathsc{V A^+A^-}|g_\mathsc{{WA^\pm B}}|^2
		\times \nonumber\\
& \times & 	
		I^{(1)}_{\sigma\mu\nu}
		(q_1,-q_2,m_\mathsc{B},m_\mathsc{A^+},m_\mathsc{A^-})
\\
		-ic_\mathsc{V} 
		{\Delta\Gamma^\mathsc{V WW}_{\sigma\mu\nu}}^{(2)}
& = &
		\left(\sum_\mathsc{THDM}-\sum_\mathsc{SM}\right)
		g_\mathsc{V W G^\pm}g_\mathsc{{WWB}}g_\mathsc{WG^\pm B}^*
		\times \nonumber \\
& \times & 	I^{(2)}_{\sigma\mu\nu}(q_1,-q_2,m_\mathsc{B},m_\mathsc{G^\pm},m_\mathsc{W})
\\
		-ic_\mathsc{V} 
		{\Delta\Gamma^\mathsc{V WW}_{\sigma\mu\nu}}^{(3)}
& = &
		\left(\sum_\mathsc{THDM}-\sum_\mathsc{SM}\right)
		\frac{1}{2}g_\mathsc{V WW}g_\mathsc{{WWB}}^2 
		\times \nonumber \\
& \times & 	I^{(3)}_{\sigma\mu\nu}(q_1,-q_2,m_\mathsc{B},m_\mathsc{W},m_\mathsc{W}) 
\\
		-ic_\mathsc{V} 
		{\Delta\Gamma^\mathsc{V WW}_{\sigma\mu\nu}}^{(4)}
& = &
		\left(\sum_\mathsc{THDM}-\sum_\mathsc{SM}\right)
		g_\mathsc{V W A^\pm B}g_\mathsc{{WA^\pm B}}^*
		\times \nonumber \\
& \times & 	I^{(4)}_{\sigma\mu\nu}(q_1,-q_2,m_\mathsc{B},m_\mathsc{A^\pm}) 
\\
		-ic_\mathsc{V} 
		{\Delta\Gamma^\mathsc{V WW}_{\sigma\mu\nu}}^{(5)}
& = &0  
\nonumber
\end{eqnarray}
Note that ${\Delta\Gamma^\mathsc{V WW}_{\sigma\mu\nu}}^{(5)}$ vanishes since it turns out to be proportional to $
		\int_0^1 {\rm d}x (1-2x)\log[m^2-x(1-x)q^2] 
		$, 
which is obviously zero. \\
Diagrams $6)$ and $7)$ provide an extra contribution to
$\Delta\Gamma^\mathsc{ZWW}$:

\begin{eqnarray}
		 -ic_\mathsc{Z}
		{\Delta\Gamma^\mathsc{ZWW}_{\sigma\mu\nu}}^{(6)}
& = &
		\left(\sum_\mathsc{THDM}-\sum_\mathsc{SM}\right) 
		g_\mathsc{ZBC}g_\mathsc{{WA^\pm B}}g_\mathsc{{WA^\pm C}}^*
		\times \nonumber \\
& \times &
		I^{(1)}_{\sigma\mu\nu}(q_1,-q_2,m_\mathsc{A^\pm},m_\mathsc{C},m_\mathsc{B}) 
\nonumber\\
		-ic_\mathsc{Z} {\Delta\Gamma^\mathsc{ZWW}_{\sigma\mu\nu}}^{(7)}& = &
		\left(\sum_\mathsc{THDM}-\sum_\mathsc{SM}\right)
		2g_\mathsc{ZZB}g_\mathsc{{WWZ}}g_\mathsc{{WWB}}
		\times \nonumber \\
& \times &
		I^{(5)}_{\sigma\mu\nu}(q_1,-q_2,m_\mathsc{W},m_\mathsc{Z},m_\mathsc{B})
\nonumber
\end{eqnarray}
The summations are taken with respect to the configurations shown in the previous table. 


\subsection{Cancellation of UV divergences \label{DivCancellation}}
\label{subsect:4:6}
Since we compute the 1-loop counterterm $\Delta\delta Z_\mathsc{TGV}$ by means of
 a specific subset of diagrams we should check that the sum of the
 UV divergences of  ${\Delta\Pi_i^\mathsc{VWW}}$ 'fit' the divergent part of
$\Delta\delta Z_\mathsc{TGV}$ to obtain a UV-finite expression for
(\ref{TGVdifference1}).

To proceed, we must first extract the divergent parts of all the ${\Delta\Gamma^\mathsc{VWW}_{\sigma\mu\nu}}^{(j)}$.
Taking into account the prescriptions specified in Appendix \ref{Integrals} we can see that the only UV-divergent integral we deal with is $I^{(1)}$. In the usual way we can isolate its divergent part of it in  the form
\[
{\rm Div}_\mathsc{UV }[I^{(1)}_{\sigma\mu\nu}(q_1,-q_2,m_j)]=-\frac{1}{24\pi^2}
{\rm C}_\mathsc{UV}T_{\sigma\mu\nu}
\]
Computing now the total UV-divergences of 
${\Delta\Gamma^\mathsc{\gamma WW}_{\sigma\mu\nu}}^{(j)}$ and 
${\Delta\Gamma^\mathsc{ZWW}_{\sigma\mu\nu}}^{(j)}$ we
get
\begin{eqnarray}
		{\rm Div}_\mathsc{UV }
		[\sum_j{\Delta\Gamma^\mathsc{\gamma WW}_{\sigma\mu\nu}}^{(j)}] 
= 
		{\rm Div}_\mathsc{UV}[{\Delta\Gamma^\mathsc{\gamma
		WW}_{\sigma\mu\nu}}^{(1)}] = &&
\nonumber\\
		-\frac{i}{e}\left(\sum_\mathsc{THDM}-\sum_\mathsc{SM}\right)
		g_\mathsc{\gamma A^+A^-}|g_\mathsc{{WA^\pm B}}|^2 
		 \frac{1}{48\pi^2}
		{\rm C}\mathsc{UV}T_{\sigma\mu\nu}
&&  \nonumber \\
=  -\frac{1}{96\pi^2}\frac{e^2}{\sin^2{\theta}}
		{\rm C}\mathsc{UV}T_{\sigma\mu\nu} &&
\end{eqnarray}
In the case of ${\Delta\Gamma^\mathsc{ZWW}}$ we have two UV divergent
topologies, namely
\begin{eqnarray}
		{\rm Div}_\mathsc{UV }[\sum_j{\Delta\Gamma^\mathsc{ZWW}_{\sigma\mu\nu}}^{(j)}]  =
		{\rm Div}_\mathsc{UV }[{\Delta\Gamma^\mathsc{ZWW}_{\sigma\mu\nu}}^{(1)}]+
		{\rm Div}_\mathsc{UV }[{\Delta\Gamma^\mathsc{ZWW}_{\sigma\mu\nu}}^{(6)}]
&& 
\nonumber\\
		= -\frac{i}{e\cot{}{\theta}}\left(\sum_\mathsc{THDM}-\sum_\mathsc{SM}\right)
		\left(g_\mathsc{\gamma A^+A^-}|g_\mathsc{{WA^\pm B}}|^2 
		+g_\mathsc{ZBC}g_\mathsc{{WA^\pm B}}g_\mathsc{{WA^\pm C}}^\dagger \right) &&
\nonumber\\
		=\frac{1}{48\pi^2}
		{\rm C}\mathsc{UV}T_{\sigma\mu\nu}=
		-\frac{1}{96\pi^2}\frac{e^2}{\sin^2{\theta}}
		{\rm C}\mathsc{UV}T_{\sigma\mu\nu} \nonumber 
\end{eqnarray}
From this we can conclude that
\begin{eqnarray}
\label{UVgammaWW}
	&& {\rm Div}_\mathsc{UV }[\Delta\Pi_{1,2,3}^\mathsc{ZWW}]  =
	{\rm Div}_\mathsc{UV }[\Delta\Pi_{1,2,3}^\mathsc{\gamma WW}]  
		 = -\frac{1}{96\pi^2}\frac{e^2}{\sin^2{\theta}} 	{\rm C}\mathsc{UV} 
\nonumber\\
	&& {\rm Div}_\mathsc{UV }[\Delta\Pi_{4,5,6,7}^\mathsc{ZWW}]  =
	{\rm Div}_\mathsc{UV }[\Delta\Pi_{4,5,6,7}^\mathsc{\gamma WW}] = 0 
\end{eqnarray}
Next, the divergent part of $\Delta\delta Z_\mathsc{TGV}$ can be easily derived from  
(\ref{CT1}):
\begin{eqnarray}
\label{counterterm_divergence_total}
		{\rm Div}_\mathsc{UV }\left[\Delta\delta Z_\mathsc{W}\right]
		 = \frac{1}{3}\left(\sum_\mathsc{THDM}-\sum_\mathsc{SM}\right)
		|g_\mathsc{WA^\pm B}|^2
		\frac{1}{16 \pi^2}
		C_\mathsc{UV} = && \nonumber \\
		= \frac{1}{96 \pi^2}\frac{e^2}
		{\sin^2{\theta_\mathsc{W}}}C_\mathsc{UV} &&
\end{eqnarray}
Comparing (\ref{UVgammaWW}) with
(\ref{counterterm_divergence_total}) we can conclude that the UV divergences in (\ref{TGVdifference1}) cancel exactly as expected.



\section{Computation of  $\Delta\Gamma^\mathsc{\gamma WW}$ 
and  $\Delta\Gamma^\mathsc{ZWW}$}
In view of the large number of relevant Feynman graphs it is 
not feasible to display all the general results in detail. 
This is mainly because of the Passarino-Veltman (PV) reduction which is traditionally used to 'scalarise' 
the tensorial structure of the resulting integrals \cite{Passarino:1978jh}  
\cite{Bardin:ak}.

Therefore we will only describe briefly some salient points, in particular the origin of the possible non-decoupling effects of heavy virtual Higgses.

\subsection{Finite part of $\Delta\delta Z_\mathsc{TGV}$}  

The mass dependence of the counterterms can be read off from (\ref{CT1}) and (\ref{CT2}). 
Assuming the 
masses of the THDM Higgs bosons to be well above $m_\mathsc{W}$  we can 
estimate the value of $\Delta\delta Z_\mathsc{W}^b$ 
to be less than about $10^{-4}$ (and falling with $m_H,m_h \to \infty$) 
i.e. small compared to the expected order of magnitude of non-decoupling effects 
($10^{-2}-10^{-3}$). 

The situation in the case of $\Delta\delta Z_\mathsc{W}^a$ is more subtle 
because of the presence of the $\mu$ scale in the logarithm in 
(\ref{CT1}). 
However, due to the above-mentioned cancellation of divergences (\ref{DivCancellation})
the overall one-loop renormalized Green functions are $\mu$-independent 
and we can either choose some particular value of $\mu$ or combine that 
term with the corresponding $\mu$-dependent factor from $\Delta\Pi^i$'s to 
obtain $\mu$-independent quantities and discuss both these pieces together. 

Note that the behaviour of the counterterms is scheme-dependent. For example in $MS$ or $\overline{MS}$ the finite parts of $\Delta\delta Z_\mathsc{W}^{a,b}$ are constant while in the on-shell scheme they typically grow logarithmically with masses of the Higgs particles in the loops. 
However, since they are strongly suppressed with respect to the finite parts of $\Delta\Pi_i$'s, this scheme-dependence is practically negligible.  

\subsection{Finite parts of $\Delta\Pi_i$'s}  
\label{subsect:5:2}
Concerning the structure of integrals contributing to $\Delta\Pi$'s 
(Appendix \ref{Integrals}), one finds out that the possible non-decoupling 
effects in the large
heavy Higgs mass regime (holding $m_\eta$ and  $m_{h^0}$ at the weak scale) 
can descend only from the divergent factors $C_{\alpha\beta\gamma}$, 
$C_{\alpha\beta}$, $B_{\alpha}$ and $B_{0}$, all the others (UV-finite) tend to 
zero.  Note that in some situations the straightforward $m_\mathsc{HEAVY}\to \infty$ limit is not meaningful because in such case the Higgs self-couplings may 
blow up and the perturbative approach used here is then no longer valid, see below.

Next, the mass-dependence of the $B$-terms seems to be much weaker compared to the highly polynomial factors in $C_{\alpha\beta\gamma}$. On the other hand, the aparent powerlike behaviour of the PV coefficients in the expansions of $C_{\alpha\beta\gamma}$'s is often compensated by the powers of heavy Higgs masses in the denominators of the PV scalar integrals $C_0$ and thus there is no reason to suppress the $B$-terms relative to the $C$-terms.  


As an illustration, consider the combination \\ 
$C_{\alpha\beta\gamma}(p_1,p_2,m_0,m_1,m_2)+sym.$ in $I^{(1)}_{\alpha\beta\gamma}$
entering the $\Delta\Pi_{1,2,3}^\mathsc{VWW}$ {\it on-shell}, i.e. taking $p_1^2=p_2^2=m_\mathsc{W}^2$, $(p_1+p_2)^2=s$.
First note that this quantity is dimensionless.
Bearing in mind how does the PV reduction work we can expect 
coefficients of three basic types (here $\tilde{B}_0$ denotes the finite part of $B_0$, $k$ is integer and $n =0,1,2\ldots$)
\begin{eqnarray}
\label{structures}
		a)\qquad &M_i^{6-n} s^k m_\mathsc{W}^{-2k+n}C_0(p^2,\ldots,M_i^2 \ldots) 
\nonumber\\
		b)\qquad &M_i^{4-n} s^k m_\mathsc{W}^{-2k+n}\tilde{B}_0(p^2,M_i^2,\ldots)
\\
		c)\qquad &M_i^{2-n} s^k m_\mathsc{W}^{-2k+n-2}
\nonumber
\end{eqnarray}
A typical contribution to $\Delta\Pi_{1,2,3}^\mathsc{VWW}$ then looks like
\[
	\Delta\Pi_{1,2,3}^\mathsc{VWW}\sim \frac{f}{16\pi^2}
|g_\mathsc{{WA^\pm B}}|^2\times X + \ldots
\]
with $X$ being an expression from the set (\ref{structures}).
Here $f$ is an $O(1)$ numerical factor, $g_\mathsc{WA^\pm B}$ are the couplings. 
Though it seems that the 
leading terms are of the order $M_i^6$, 
 such a growth is in fact reduced by factors involving negative powers of masses, 
coming from denominators of the $C_0$ functions in the heavy Higgs mass regime. 
The suppression is even stronger once the parameters obey the decoupling limit behaviour, see below. 
Therefore one has to be very careful in semiquantitative arguments based on (\ref{structures}). 

Unfortunately, due to the enormous complexity of the results it is almost hopeless to try to get simple general analytic expressions for the leading terms in $\Delta\Pi_{i}^\mathsc{VWW}$'s even in the heavy Higgs mass regime. 
Next, it is worth {\it focusing namely to the cases when one can not expect the decoupling of the additional Higgs bosons in the Appelquist-Carazzone manner}.
   
For such setups we have performed at least a simple numerical analysis with the following results\footnote{We have also checked the proper decoupling behaviour of the formfactors in the cases of having the heavy Higgs sector adjusted towards the decoupling regime; this provides a simple consistency check of the results.}:
\\
\\
1) The leading terms in $|\Delta\Pi_i^\mathsc{VWW}|$'s 
typically contain logarithms and inverse-powers of the heavy Higgs masses, i.e.  
\[
	|\Delta\Pi_i^\mathsc{VWW}|\sim \frac{f}{16\pi^2}
|g_1 g_2^*|\left[k_1\,\log \frac{m_\mathsc{HIGGS}}{m_\mathsc{W}} + 
k_2\, O(m_\mathsc{W}m_\mathsc{HIGGS}^{-1})\right]+\ldots
\]
This means that the larger the Higgs masses are, the smoother $\Delta\Pi_i^\mathsc{VWW}$ behave and their behaviour tends to be purely logarithmic (of course, with $k_1\to 0$ in the decoupling regime).
\\
2) The overall magnitudes of  $|\Delta\Pi_i^\mathsc{VWW}|$ usually turn out to be around $10^{-3}$ (at $m_\mathsc{HIGGS} \sim m_\mathsc{W}$) so the possible large non-decoupling effects in physical amplitudes seem to be quite unlikely, barring some special enhancements coming from kinematics and/or geometrical factors \cite{MalinskyHorejsi2}. 
\\ 
 Let us illustrate these features in the case of $|\,\,\Delta\Pi_1^\mathsc{VWW}|$ 
and $|\,\,\Delta\Pi_2^\mathsc{VWW}|$ within one of the concrete realisations
 of the Higgs sector of the model. Although the Higgs self-couplings do not 
enter explicitly  our analysis, their values determine the shape of the Higgs 
spectrum of the model which must be chosen in a way 
compatible with these constraints. In other words, shifting the heavy Higgs
 masses and holding at the same time some of the features of the Higgs potential
 unchanged causes a shift in the mixing angles $\alpha$ and $\beta$ which
 propagates via $\sin (\alpha-\beta)$ and $\cos (\alpha-\beta)$ to the 
vector-boson-Higgs couplings. 
\\
For simplicity, we take 
\footnote{N.B. It is well known that the choice (\ref{setup}) provides a setup in which the heavy Higgs mass limit does not 
exist at all \cite{Gunion:2002zf}, i.e. we can not push the Higgs masses too far 
beyond the weak scale. This is caused by the fact that in this setup all 
the THDM Higgs masses must be of the same order:
\[
m_h^2+m_H^2 = \lambda v^2 \quad m_A^2 = -\Re\lambda_5 v^2 \quad m^2_{H^\pm}=-\frac{1}{2}(\lambda_4+\Re\lambda_5)v^2
\]
Then the only way to get large masses of $H$, $A$ and $H^\pm$ consists in 
having $\lambda$, $|\lambda_4|$ and ${\rm Re}\,\lambda_5$ well above 1. 
In other words, it is exactly one of the physically interesting situations 
in which one can expect non-decoupling behaviour of the heavy part of the 
THDM Higgs spectrum and some (in principle) measurable deviations from the SM predictions; 
for further details see for example \cite{MalinskyHorejsi2}. 

The main advantage of the choice is the simplicity 
of the relation for $\cos^2 (\alpha-\beta)$ which is the basic ingredient 
of any quantitative analysis.

}
\begin{equation}\label{setup}
\lambda_6=\lambda_7=0, \qquad m_{12}=0
\end{equation}
and especially 
\[
\lambda_1=\lambda_2\equiv \lambda \qquad \lambda_3=1, \quad {\rm and} \quad \beta=\pi/2
\nonumber
\]
The remaining parameters $\lambda$, $\lambda_4$, $\lambda_5$ $\alpha$, $m_1$ and $m_2$ are then driven by the choice of $m_h$, $m_H$, $m_A$, $m_{H^\pm}$ so that the resulting Higgs potential produces the right mass pattern.\\\mbox{}\\ 
For example, let us display the behaviour of the $|\Delta\Pi_{1}^\mathsc{\gamma WW}|$ and $|\Delta\Pi_{1}^\mathsc{\gamma WW}|$ in this model as functions of the $m_{A^0}$ parameter. The other parameters are fixed as follows: $m_\eta = 105$GeV, $m_{h^0} = 125$GeV, $m_{H^0} = 145$GeV, $m_{H^\pm} = 180$GeV, $\sqrt{s}=250$GeV.  
\begin{center}
\epsfig{file=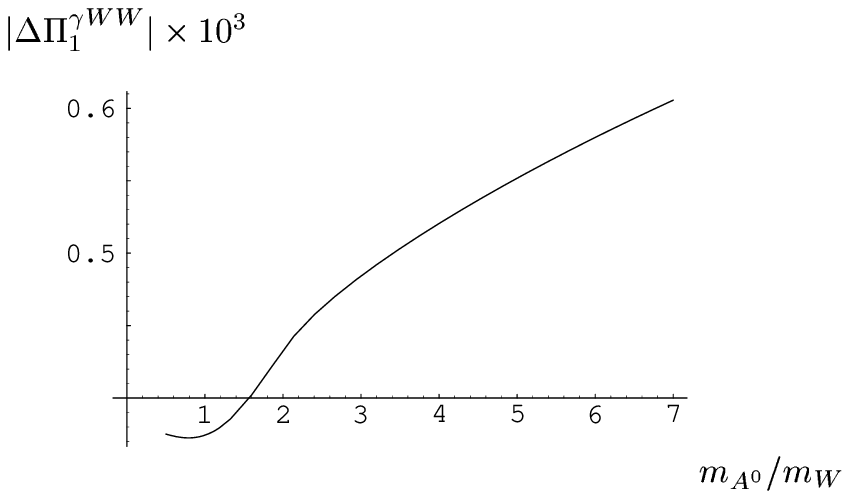,width=8cm}
\\ 
\epsfig{file=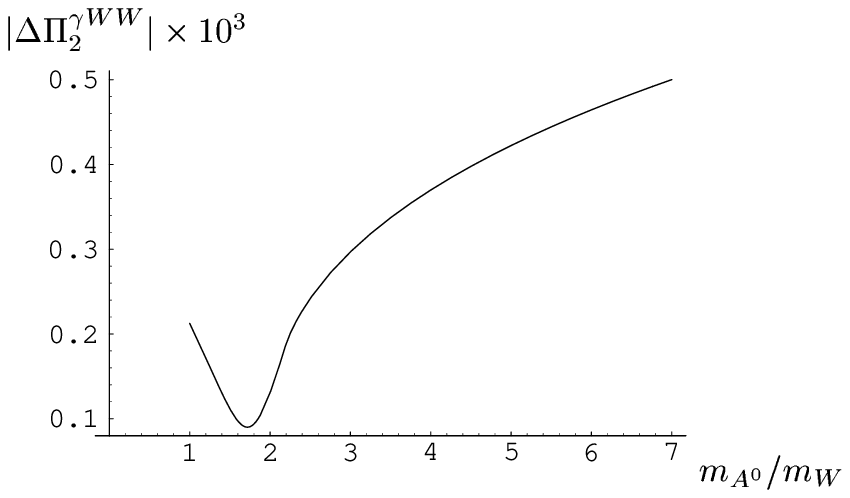,width=8cm}
\end{center}
As was stated above,  these quantities grow logarithmically (in the non-decoupling setup (\ref{setup}) ) with the mass of the relatively heavy $m_A^0$ boson in the model.



\section{Conclusion}
We have  computed the additional contributions to the one-loop 
THDM triple gauge vertices which are not present within the SM framework.
The model is taken to be very general \cite{Gunion:1989we} 
with no need of any additional constraints to its structure. 
\\
We have adopted the on-shell renormalization scheme, using the 
dimensional regularization of UV divergences.
The finite parts of the on-shell counterterms are 
computed with the help of the $W$-boson propagator renormalization constants. Cancellation of UV 
divergences is checked explicitly in both $\gamma WW$ and $ZWW$ cases. 
\\
The THDM heavy Higgs boson 
contributions to the triple gauge vertices can in some situations lead to 
possible non-decoupling effects in physical amplitudes. 
Therefore, these results can be employed (at least in principle) 
for an indirect exploration of the structure of the electroweak 
Higgs sector at future collider facilities.


\section*{Acknowledgements}
This work was supported by "Centre for Particle Physics", project No. LN00A006 of the 
Ministry of Education of the Czech Republic.



\appendix

\section{THDM interactions of vector bosons}
\label{AppendixA}
\paragraph{Relevant part of THDM lagrangian}\mbox{}\\
The lagrangian (\ref{lagrangian}) can be decomposed into three parts corresponding to VHH, VVH and VVHH vertices respectively:
$$ 
{\cal L}^\mathsc{KIN.}_\mathsc{HIGGS}=
		= {\cal L}_\mathsc{VVH}+{\cal L}_\mathsc{VHH}+{\cal L}_\mathsc{VVHH}+
		{\rm other \,\,\, terms}.
$$
Here
\begin{eqnarray}
		{\cal L}_\mathsc{VVH} 
& = & 
		e m_\mathsc{Z}\cot{\theta_\mathsc{W}} {W^+}^\mu W^-_\mu 	
		\left[H^0\cos(\alpha-\beta)- h^0\sin(\alpha-\beta)\right] 
\nonumber \\
& + &
		\frac{e m_\mathsc{Z}}{\sin{2\theta_\mathsc{W}}} {Z}^\mu Z_\mu
		\left[H^0\cos(\alpha-\beta)- h^0\sin(\alpha-\beta)\right]+  
\nonumber \\
& + &
		e m_\mathsc{Z}\left(\cos{\theta_\mathsc{W}} {A}^\mu  - 
		\sin{\theta_\mathsc{W}} Z^\mu\right) \left(W^+_\mu G^- + W^-_\mu G^+\right) \nonumber
\end{eqnarray}
and
\begin{eqnarray}
&& {\cal L}_\mathsc{VHH} = i e A^\mu H^+\partial_\mu^\leftrightarrow H^-
		+i e A^\mu  G^+ \partial_\mu^\leftrightarrow G^- +
\nonumber \\
&& +   
		i e \cot{{2\theta_{\mathsc{W}}}} Z^\mu H^+\partial_\mu^\leftrightarrow H^-
		+i e \cot{{2\theta_{\mathsc{W}}}} Z^\mu  G^+ \partial_\mu^\leftrightarrow G^-
+ \nonumber \\
&& + 
		\frac{e}{\sin{{2\theta_{\mathsc{W}}}}}\times \nonumber \\ 
&& \times \left\{  Z^\mu \left[
		\cos(\alpha-\beta) A^0 \partial_\mu^\leftrightarrow h^0-\sin(\alpha-\beta) G^0 				\partial_\mu^\leftrightarrow h^0
		\right]\right.
+ \nonumber \\
&& + 
		  Z^\mu \left[
		\sin(\alpha-\beta) A^0 \partial_\mu^\leftrightarrow H^0+\cos(\alpha-\beta) G^0 				\partial_\mu^\leftrightarrow H^0
		\right] 
- \nonumber \\
&& -
		\cos{(\alpha-\beta)}
		\left( {W^-}^\mu h^0 \partial^\leftrightarrow_\mu H^+ + {W^+}^\mu H^- 		
		\partial^\leftrightarrow_\mu
		h^0\right)
+ \nonumber \\
&& +
		  \sin{(\alpha-\beta)}
		\left( {W^-}^\mu h^0 \partial^\leftrightarrow_\mu G^+ + {W^+}^\mu G^-
		\partial^\leftrightarrow_\mu
		h^0\right) 
- \nonumber \\
&& -
		 \sin{(\alpha-\beta)}
		\left( {W^-}^\mu H^0 \partial^\leftrightarrow_\mu H^+ + {W^+}^\mu H^-
		\partial^\leftrightarrow_\mu
		H^0\right) 
- \nonumber \\
&& -
		 \cos{(\alpha-\beta)}
		\left( {W^-}^\mu H^0 \partial^\leftrightarrow_\mu G^+ + {W^+}^\mu G^- 
		\partial^\leftrightarrow_\mu
		H^0\right) 
- \nonumber \\
&& -
		  \left({W^-}^\mu A^0	
		\partial_\mu^\leftrightarrow H^+ - {W^+}^\mu H^- \partial_\mu^\leftrightarrow A^0
		\right) 
- \nonumber \\
&& -
		 \left. \left({W^-}^\mu G^0 	
		\partial_\mu^\leftrightarrow G^+ - {W^+}^\mu G^- \partial_\mu^\leftrightarrow G^0
		\right)\right\}
\nonumber 
\end{eqnarray}
while
\begin{eqnarray}
&&		{\cal L}_\mathsc{VVHH}
		= \frac{e^2}{2\sin^2{\theta_{\mathsc{W}}}} {W^+}^\mu W^-_\mu \left[H^+H^- + G^+G^- 
		\right]+  
\nonumber \\
&& +
		\frac{e^2}{4\sin^2{\theta_{\mathsc{W}}}} {W^+}^\mu W^-_\mu
		\left[(h^0)^2+(H^0)^2+(A^0)^2+(G^0)^2\right]+
\nonumber \\
&& + 
		\frac{e^2}{2\sin{\theta_{\mathsc{W}}}} A^\mu
		\left[H^0 \cos(\alpha-\beta)-h^0 \sin(\alpha-\beta)\right]
\times \nonumber \\ && \hskip 2cm  \times 
		\left(W^+_\mu G^- + W^-_\mu G^+\right)
		+
\nonumber \\
&& +
		\frac{e^2}{2\sin{\theta_{\mathsc{W}}}} A^\mu
		\left[H^0 \sin(\alpha-\beta)+h^0 \cos(\alpha-\beta)\right]
\times \nonumber \\ &&  \hskip 2cm \times 
		\left(W^+_\mu H^- + W^-_\mu H^+\right)-
\nonumber \\
&& - 
		\frac{e^2}{2\cos{\theta_{\mathsc{W}}}} Z^\mu
		\left[H^0 \cos(\alpha-\beta)-h^0 \sin(\alpha-\beta)\right]
\times \nonumber \\ && \hskip 2cm  \times 
		\left(W^+_\mu G^- + W^-_\mu G^+\right)+
\nonumber \\
&& -
		\frac{e^2}{2\cos{\theta_{\mathsc{W}}}} Z^\mu
		\left[H^0 \sin(\alpha-\beta)+h^0 \cos(\alpha-\beta)\right]
\times \nonumber \\ && \hskip 2cm  \times 
		\left(W^+_\mu H^- + W^-_\mu H^+\right)+
\nonumber \\
&& + 
		\frac{ie^2}{\sin{2\theta_{\mathsc{W}}}} G^0 \left(\cos{\theta_{\mathsc{W}}} A^\mu
		- \sin{\theta_{\mathsc{W}}} Z^\mu  \right)
\times \nonumber \\ && \hskip 2cm \times 
		\left(W^+_\mu G^- - W^-_\mu G^+\right)+
\nonumber \\
&& +
		\frac{ie^2}{\sin{2\theta_{\mathsc{W}}}} A^0 \left(\cos{\theta_{\mathsc{W}}} A^\mu
		- \sin{\theta_{\mathsc{W}}} Z^\mu  \right)
\times \nonumber \\ && \hskip 2cm \times 
		\left(W^+_\mu H^- - W^-_\mu H^+\right)
\nonumber 
\end{eqnarray}

\paragraph{Coupling constants\label{couplings}}\mbox{}\\
As before, the charged (pseudo-) scalars are denoted by the generic symbols $A^\pm$ while the neutral by $B$ and $C$.
\paragraph{VVHH type:}
\begin{center}
    \renewcommand{\arraystretch}{1.5}
    \begin{tabular}{c|cc}
    ${ A^\pm B}$    & ${  g_\mathsc{\gamma WA^\pm B}}$& ${  g_\mathsc{ZWA^\pm B}}$\\
    \hline
$ G^\pm\eta$ & $ \frac{1}{2}ie^2\cs{}{\theta}$& $
-\frac{1}{2}ie^2\sc{}{\theta}$ \\
$ G^\pm h^0 $ & $ -\frac{1}{2} i e^2
\cs{}{\theta} \s{}{\alpha-\beta} $ & $ \frac{1}{2} i e^2 \sc{}{\theta}
\s{}{\alpha-\beta} $\\
$ G^\pm H^0$ & $ \frac{1}{2} i e^2 \cs{}{\theta}
\c{}{\alpha-\beta} $ & $ -\frac{1}{2} i e^2 \sc{}{\theta} \c{}{\alpha-\beta} $
\\
$ H^\pm h^0$ &  $ \frac{1}{2} i e^2 \cs{}{\theta} \c{}{\alpha-\beta}$ &  $
-\frac{1}{2} i e^2 \sc{}{\theta} \c{}{\alpha-\beta}$  \\
$ H^\pm H^0$ & $
\frac{1}{2} i e^2 \cs{}{\theta} \s{}{\alpha-\beta}$ & $ -\frac{1}{2} i e^2
\sc{}{\theta} \s{}{\alpha-\beta}$\\
$ G^\pm G^0$ & $ -\frac{1}{2}e^2
\cs{}{\theta}$ & $ \frac{1}{2}e^2 \sc{}{\theta}$ \\
$ H^\pm A^0$ & $
-\frac{1}{2}e^2 \cs{}{\theta}$ & $ \frac{1}{2}e^2 \sc{}{\theta}$
\end{tabular} 
\end{center}

\paragraph{VHH type:}
\begin{center}
\renewcommand{\arraystretch}{1.5}
    \begin{tabular}{c|c}
    ${ BC}$ & ${  g_\mathsc{ZBC}}$\\
    \hline
    $G^0\eta $ & $ -e   \cs{}{2\theta} $  \\
    $G^0h^0 $ & $ e    \cs{}{2\theta} \s{}{\alpha-\beta}$ \\
    $G^0H^0 $ & $ -e    \cs{}{2\theta} \c{}{\alpha-\beta}$  \\
    $A^0h^0 $ & $ -e    \cs{}{2\theta} \c{}{\alpha-\beta}$  \\
    $A^0H^0 $ & $ -e    \cs{}{2\theta} \s{}{\alpha-\beta}$
  \end{tabular}
  \hskip 1cm
    \begin{tabular}{c|c}
     ${ A^\pm B}$   & ${  g_\mathsc{WA^\pm B}}$ \\
    \hline
 $ G^\pm\eta$ & $ \frac{1}{2}ie   \cs{}{\theta}$ \\
 $ G^\pm h^0 $ & $ -\frac{1}{2}ie    \cs{}{\theta} \s{}{\alpha-\beta} $ \\
 $ G^\pm H^0$ & $ \frac{1}{2}ie    \cs{}{\theta} \c{}{\alpha-\beta} $  \\
 $ H^\pm h^0$ &  $ \frac{1}{2}ie    \cs{}{\theta} \c{}{\alpha-\beta}$ \\
 $ H^\pm H^0$ & $ \frac{1}{2}ie    \cs{}{\theta} \s{}{\alpha-\beta}$\\
 $ G^\pm G^0$ & $ -\frac{1}{2}e\cs{}{\theta}$ \\
 $ H^\pm A^0$ & $ -\frac{1}{2}e\cs{}{\theta}$
  \end{tabular}
    \begin{tabular}{c|cc}
    ${A^+A^-}$  & ${  g_\mathsc{\gamma A^+A^-}}$&  ${  g_\mathsc{ZA^+A^-}}$ \\
    \hline
    $G^+G^- $ & $ -ie $ & $ -ie   \ct{}{2\theta}$ \\
    $H^+H^- $ & $ -ie $ & $ -ie   \ct{}{2\theta}$
    \end{tabular}
\end{center}

\paragraph{VVH type:}
\begin{center}
\renewcommand{\arraystretch}{1.5}
\begin{tabular}{c|cc}
    $B$ & ${  g_\mathsc{ZZB}}$ & ${  g_\mathsc{WWB}}$\\
    \hline
    $\eta$ & $ie m_\mathsc{Z}   \cs{}{2\theta}$ & $ie m_\mathsc{Z}  \ct{}{\theta}$ \\
    $h^0$ & $-ie m_\mathsc{Z}   \cs{}{2\theta} \s{}{\alpha-\beta}$ & $-ie m_\mathsc{Z}  \ct{}{\theta} \s{}{\alpha-\beta}$ \\
    $H^0$ & $ie m_\mathsc{Z}   \cs{}{2\theta} \c{}{\alpha-\beta}$ & $ie m_\mathsc{Z}  \ct{}{\theta} \c{}{\alpha-\beta}$
\end{tabular}
\hskip 1cm
$
\begin{array}{ll}   
    g_\mathsc{\gamma W G^\pm}= iem_\mathsc{Z} \c{}{\theta} \\
    g_\mathsc{ZW G^\pm}= -iem_\mathsc{Z} \s{}{\theta}
\end{array}
$
\end{center}

\paragraph{VVV type:}
\begin{displaymath}
    g_\mathsc{\gamma WW}= -ie \qquad \qquad
  g_\mathsc{Z WW}= -ie\ct{}{\theta}
\end{displaymath}


\section{Useful integrals}
\label{Integrals}
At this place we display all necessary loop integrals in terms of the Passarino-Veltman functions \cite{Bohm:1987ck},\cite{Passarino:1978jh},\cite{Bardin:ak} listed below.
\begin{eqnarray}
&&		I^{(1)}_{\alpha\beta\gamma}(p_1,p_2,m_0,m_1,m_2)\equiv 
		\left(\begin{array}[pos]{c}
		\beta \leftrightarrow \gamma \\
		p_1 \leftrightarrow p_2
		\end{array}\right)+ \mu^{2\varepsilon}i^3 \times
\nonumber\\
&& 
		\times
		\int\frac{d^dk}{(2\pi)^d}
		\frac{-(2k+p_1+p_2)_\alpha(2k+p_1)_\beta(2k+p_2)_\gamma}
		{(k^2-m_0^2)[(k+p_1)^2-m_{1}^2][(k+p_2)^2-m_{2}^2]}=
\nonumber\\
&& -
		\frac{1}{16\pi^2}\left[8C_{\alpha\beta\gamma}+
		4(p_1+p_2)_\alpha C_{\beta\gamma}+
		4{p_1}_\beta C_{\alpha\gamma}+
		4{p_2}_\gamma C_{\alpha\beta} \nonumber \right. \\
&&
		+2(p_1+p_2)_\alpha {p_1}_\beta C_{\gamma}+ 
		2(p_1+p_2)_\alpha {p_2}_\gamma C_{\beta}+
		2{p_1}_\beta {p_2}_\gamma C_{\alpha}+
\nonumber\\
&& + 
		\left.
		(p_1+p_2)_\alpha {p_1}_\beta {p_2}_\gamma 	
		C_0\right](p_1,p_2,m_0,m_1,m_2)
\nonumber\\
&& +
		(p_1 \leftrightarrow p_2, \beta \leftrightarrow \gamma)
\nonumber 
\end{eqnarray}
\begin{eqnarray}
&&		I^{(2)}_{\alpha\beta\gamma}(p_1,p_2,m_0,m_1,m_2)\equiv 
		\left(\begin{array}[pos]{c}
		\beta \leftrightarrow \gamma \\
		p_1 \leftrightarrow p_2
		\end{array}\right) + \mu^{2\varepsilon}i \times
\nonumber\\
&&		\times
		\int\frac{d^dk}{(2\pi)^d}
		\frac{g_{\alpha\gamma}(2k+p_1)_\beta}
		{(k^2-m_0^2)[(k+p_1)^2-m_{1}^2][(k+p_2)^2-m_{2}^2]}
=
\nonumber\\
&& -
		\frac{1}{16\pi^2} g_{\alpha\gamma}\left[
		2 C_{\beta}+{p_1}_\beta C_0\right](p_1,p_2,m_0,m_1,m_2)+
\nonumber\\
&& +(p_1 \leftrightarrow p_2, 		\beta \leftrightarrow \gamma)
		\label{integrals}
\end{eqnarray}
\begin{eqnarray}
&&		I^{(3)}_{\alpha\beta\gamma}(p_1,p_2,m_0,m_1,m_2)\equiv
		(p_1 \leftrightarrow p_2, \beta \leftrightarrow \gamma)
		+\mu^{2\varepsilon} i^3\times
\nonumber\\
&&
		\!\!
		\int
		\!\!
		\frac{d^dk}{(2\pi)^d}
		\frac{g_{\beta\gamma}(2k\!\!+\!\!p_1\!\!+\!\!p_2)_\alpha
		\!-\!g_{\alpha\gamma}(2p_2\!\!-\!\!p_1\!\!+\!\!k)_\beta
		\!-\!g_{\alpha\beta}(2p_1\!\!-\!\!p_2\!\!+\!\!k)_\gamma}
		{(k^2-m_0^2)[(k+p_1)^2-m_{1}^2][(k+p_2)^2-m_{2}^2]}
\nonumber\\
&&=
		\frac{1}{16\pi^2}\left\{2g_{\beta\gamma}C_\alpha-g_{\alpha\gamma}C_\beta-
		g_{\alpha\beta}C_\gamma+
		\left[g_{\beta\gamma}(p_1+p_2)_\alpha-\right. \right. 
\nonumber\\		
&&
		\left.\left.
		\!-\!g_{\alpha\gamma}(2p_2\!-\!p_1)_\beta
		\!-\!g_{\alpha\beta}(2p_1\!-\!p_2)_\gamma\right]C_0
		\right\}(p_1\!,p_2\!,m_0\!,m_1\!,m_2) 
\nonumber\\
&&+
		(p_1 \leftrightarrow p_2, \beta \leftrightarrow \gamma)
\nonumber 
\end{eqnarray}
\begin{eqnarray}
&&		I^{(4)}_{\alpha\beta\gamma}(p_1,p_2,m_0,m_1)\equiv 
		\mu^{2\varepsilon} i^2 \times
\nonumber\\
&&
		\times
		\int\frac{d^dk}{(2\pi)^d}
		\frac{g_{\alpha\gamma}(2k+p_1)_\beta}
		{(k^2-m_0^2)[(k+p_1)^2-m_{1}^2]}+\left(\begin{array}[pos]{c}
		\beta \leftrightarrow \gamma \\
		p_1 \leftrightarrow p_2
		\end{array}\right) =
\nonumber\\
&&=
		-\frac{i}{16\pi^2} g_{\alpha\gamma}\left[
		2 B_{\beta}\!+\!{p_1}_\beta B_0\right](p_1\!,m_0\!,m_1)
		+(p_1 \!\leftrightarrow\! p_2, \beta \!\leftrightarrow\! \gamma)
\nonumber 
\end{eqnarray}
\begin{eqnarray}
&&
		I^{(5)}_{\alpha\beta\gamma}(p_1,p_2,m_0,m_1,m_2)\equiv  
		\left(\begin{array}[pos]{c}
		\beta \leftrightarrow \gamma 
		\\
		p_1 \leftrightarrow p_2
		\end{array}\right) + \mu^{2\varepsilon} i^3 \times
\nonumber\\
&&
	        \int\frac{d^dk}{(2\pi)^d}
		\frac{g_{\beta\gamma}(p_1-k)_\alpha+g_{\alpha\gamma}(2k+p_1)_\beta
		-g_{\alpha\beta}(2p_1+k)_\gamma}
		{(k^2-m_0^2)[(k+p_1)^2-m_{1}^2][(k+p_2)^2-m_{2}^2]}
\nonumber\\
&&
=
		\frac{1}{16\pi^2}
		\left\{
		-g_{\beta\gamma}C_\alpha
		+2g_{\alpha\gamma}C_\beta
		-g_{\alpha\beta}C_\gamma+
		\right. 
\nonumber\\
&&
+
		\left.
		\left[g_{\beta\gamma}{p_1}_\alpha\!+\!g_{\alpha\gamma}{p_1}_\beta
		\!-\!g_{\alpha\beta}{2p_1}_\gamma\right]C_0
		\right\}
		(p_1\!,p_2\!,m_0\!,m_1\!,m_2)+
\nonumber\\
&&+		
		(p_1 \leftrightarrow p_2, \beta \leftrightarrow \gamma)
\nonumber
\end{eqnarray}
The Passarino-Veltman functions are defined as
\begin{eqnarray}
&&		\frac{i}{16\pi^2}
		B_0,B_\alpha,B_{\alpha\beta}(p,m_0,m_1)\equiv 
\nonumber \\
&& \equiv \mu^{2\varepsilon}
		\int\frac{d^dk}{(2\pi)^d}
		\frac{1,k_\alpha, k_{\alpha\beta}}
		{(k^2-m_0^2)[(k+p)^2-m_{1}^2]}
\nonumber \\ 
&&		\frac{i}{16\pi^2}
		C_0,C_\alpha,C_{\alpha\beta},C_{\alpha\beta\gamma}(p_1,p_2,m_0,
		m _ 1 , m _ 2  )\equiv 
\nonumber\\
&& \equiv 
		\mu^{2\varepsilon}
		\int\frac{d^dk}{(2\pi)^d}
		\frac{1, k_\alpha,k_\alpha k_\beta, k_\alpha k_\beta k_\gamma}
		{(k^2-m_0^2)[(k+p_1)^2-m_{1}^2][(k+p_2)^2-m_{2}^2]}
\nonumber
\end{eqnarray}
(calculations are performed in dimension $d=4-2\varepsilon$, propagators are displayed without the '$+i\eta$' factors
)





%

\end{document}